\documentclass[9pt,twocolumn,twoside]{optica}
\setboolean{shortarticle}{false}
\setboolean{minireview}{false}

\usepackage{graphicx}
\usepackage{multicol}
\usepackage{multirow}
\usepackage{tabularx}
\usepackage{lscape}
 \newcolumntype{b}{>{\centering\arraybackslash}>{\hsize=2.3\hsize}X}
\newcolumntype{s}{>{\centering\arraybackslash}>{\hsize=.45\hsize}X}
\newcolumntype{m}{>{\centering\arraybackslash}>{\hsize=.9\hsize}X}
    
\usepackage{array, makecell} %
\usepackage{amsmath}
\usepackage[normalem]{ulem}
\usepackage{dblfloatfix} 

\title{Exploiting wavelength diversity for high resolution time-of-flight 3D imaging}

\author[1,$\dag$,*]{Fengqiang Li}
\author[2,$\dag$]{Florian Willomitzer}
\author[3]{Prasanna Rangarajan}
\author[1]{Oliver Cossairt}

\affil[$\dag$]{These authors are considered joint first authors}
\affil[1]{Department of Computer Science, Northwestern University, Evanston, IL, 60208, USA}
\affil[2]{Department of Electrical and Computer Engineering, Northwestern University, Evanston, IL, 60208, USA}
\affil[3]{Department of Electrical and Computer Engineering, Southern Methodist University, Dallas, TX, 75205, USA}
\affil[*]{Corresponding author: fengqiang.li@u.northwestern.edu}

\begin{abstract}
State-of-the-art time-of-flight (ToF) based 3D sensors suffer from poor lateral and depth resolutions. In this work, we introduce a novel sensor concept that provides ToF-based 3D measurements of real world objects with depth precisions up to 35$\mu m$ and point cloud densities at the native sensor-resolutions of state-of-the-art CMOS/CCD cameras (up to several megapixels). Unlike other continuous-wave amplitude-modulated ToF principles, our approach exploits wavelength diversity for an interferometric surface measurement of macroscopic objects with rough or specular surfaces. Based on this principle, we introduce three different embodiments of prototype sensors, exploiting three different sensor architectures.

\end{abstract}
\setboolean{displaycopyright}{false}

\begin{document}

\maketitle

\section{Introduction}

\label{sec:intro}
Recent years have witnessed an explosion of interest in three-dimensional (3D) imaging and sensing technologies~\cite{lidarSelfDriving,foix2011lock,lazaros2008review,levin2007image,schechner2000depth,antipa2018diffusercam,morimoto2020megapixel,eigen2014depth}. Common applications include medical 3D imaging \cite{Dalal14, Huber11,huang1991optical,li2014label},  precision engineering, industrial inspection~\cite{molleda2013improved,lamoreux2004space,Caulier:10}, absolute distance metrology~\cite{liu2011sub,piracha2010range, Willo15}, autonomous navigation~\cite{lidarSelfDriving,hee2013motion}, documentation of cultural heritage \cite{Willo20PMD, yeh2016streamlined}, and virtual/augmented reality~\cite{hololens,izadi2011kinectfusion, Willo3DCam17}.

The primary objective of any 3D image sensing modality is to provide a spatially resolved 3D representation of the imager field of view (FoV). The ability to localize targets in three dimensions distinguishes 3D sensing from traditional imaging. A survey of literature in 3D sensing reveals an abundance of sensing principles that can be broadly categorized into three distinct groups: (1) Triangulation-based approaches (including active and passive stereo), (2) Time-of-Flight ("ToF")-based approaches (including multi/single wavelength interferometry, LIDAR, and so-called "ToF-cameras"), and (3) Reflectance-based approaches that measure the surface gradient (including photometric stereo and deflectometry). In the following we concentrate on \textit{active} measurement principles whose intrinsic measurand is the 3D \textit{shape} (triangulation and ToF-based methods), and not the \textit{slope} (reflectance-based methods). We briefly outline the related measurement principles and the associated limits.

\textbf{Active triangulation based approaches} spatially modulate the incident irradiance and exploit the difference in the viewpoint of the source and sensor to recover 3D information using triangulation principles. These methods provide high depth precision (down to 100s of $\mu m$) and high lateral resolution \cite{Harendt14,WilloOSAV13}, making it possible to record densely sampled representations of a 3D scene. Recent advances have additionally demonstrated the ability to capture high resolution 3D images at video rates \cite{Schaffer10, Heist16,Willo3DCam17, GerdWhy}. 

\begin{figure}[htbp]
    \includegraphics[width=1\linewidth]{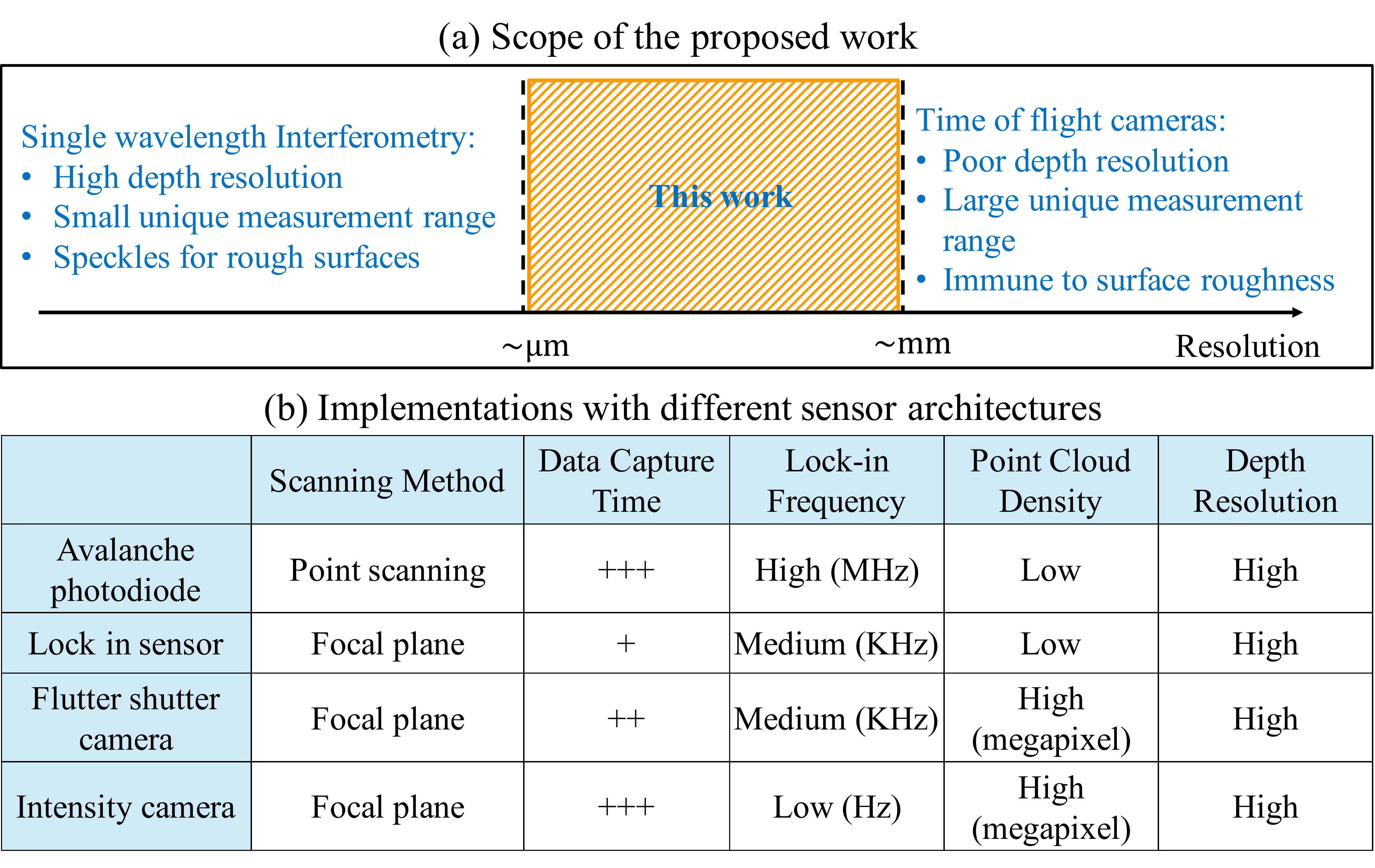}
    \caption{(a) Scope of this work: Our focal-plane array (FPA) based 3D sensors provide a niche between optical interferometry and ToF cameras in terms of performance, while also avoiding the need for raster scanning~\cite{li2018sh}. (b) We implement three different FPA-based architectures exploiting different data capture speed, spatial data density, and lock-in frequency.}
    \label{fig:scope}
\end{figure}

The chief limitations of triangulation based 3D sensors include difficulty in accommodating specular or transparent object surfaces. Moreover, the depth precision $\delta z$ of triangulation based schemes varies directly with the baseline and inversely with standoff distance. As a result, large baselines are required to realize higher depth precision at longer standoff distances, and the geometrical arrangement of the sensor components can lead to occlusion problems during the measurement.

\textbf{ToF based approaches} recover depth information by exploiting a temporally modulated illumination that is correlated with a reference signal. The principal advantage of ToF approaches over triangulation based approaches to 3D sensing is the elimination of the baseline requirement, not at least because the depth precision is independent of the standoff distance. These features make ToF approaches particularly attractive for depth sensing in compact form factors.

For the so-called \textit{\textbf{"ToF sensors"}} the temporal modulation of the signal is realized by modulating the light source radiance.  Common choices include: pulsed waveform modulation~\cite{weitkamp2006lidar} and continuous wave amplitude modulation (CWAM)~\cite{foix2011lock}. Pulsed approaches to ToF sensors such as LIDAR can provide depth precisions in the $mm$ range. However, most LIDAR systems measure only one 3D point at a time, and rely on raster scanning to produce a full 3D image, which limits the acquisition speeds that can be achieved. Furthermore, high precision depth measurement demands extensive calibration of the scanning mechanism and electronic synchronization of the illumination and sensing subsystems. CWAM based ToF sensors (so-called "ToF cameras") exploit sinusoidal amplitude modulation ~\cite{lange2001solid}. Each detector pixel in such a ToF camera demodulates the received irradiance to extract the phase shift between signal and reference to subsequently calculate the target depth. It is worth emphasizing that a ToF pixel operates unlike a traditional pixel which merely integrates the received irradiance. This distinction allows ToF cameras to simultaneously record the brightness of a target and its depth.  Commercially available ToF cameras~\cite{schwarte1997new,kinect_sensor,TIToF} are able to capture 3D images at video frame rates. These devices are being widely used for computer vision (CV) tasks such as object recognition and tracking in gaming and virtual/augmented reality (VR/AR)~\cite{hololens}. Such state-of-the-art ToF cameras can achieve a depth precision of centimeters~\cite{kinect_sensor}. This resolution, while adequate for common CV and VR/AR tasks, is inadequate for precision measurement tasks such as industrial metrology.

To better appreciate the origins of the limited depth precision of ToF based techniques, we have to more closely examine their operating principle: At each pixel, the depth $z$ to the target may be estimated from the measured phase shift $\phi_{t}$ between measured signal and reference as $z = c\phi_{t}/(2 \pi f_{t})$, where $f_{t}$ is the modulation frequency of the light source, and $c$ is the speed of light. The depth precision $\delta z$ is inversely proportional to the modulation frequency~\cite{born2013principles, lange2001solid,kadambi2017rethinking}: $\delta z \sim 1/f_{t}$. 

In a commercial ToF camera, each pixel behaves as a homodyne receiver~\cite{lange2001solid, foix2011lock} that accumulates a charge proportional to $\phi_{t}$. These cameras use modulation frequencies in tens of MHz. In principle, higher modulation frequency (order of GHz) can be used to achieve submillimeter depth precision. However, current manufacturing technology limits the maximum modulation frequencies that can be achieved in silicon~\cite{kinect_sensor,TIToF}. 

At the opposite end of the spectrum, we have \textbf{\textit{Interferometric ToF-approaches}} which exploit the modulation intrinsic to the process of recording the interference of the target return with a coherent reference. The resulting optical modulation frequencies are 100's of THz (instead of electronic frequencies around MHz). As a result, optical interferometry ~\cite{born2013principles,michelson1881relative} can easily achieve sub-$\mu m$ depth precision. The improvement, however, comes at the expense of increased sensitivity to environmental fluctuations and target surface roughness which produce speckles~\cite{goodman1975statistical}.  The authors of~\cite{dandliker1988two, fercher1985rough} demonstrated that the measurement at a second wavelength compensates for these problems. The idea was further extended in~\cite{li2017high,li2018sh} by providing a flexible trade-off between unique measurement range and depth precision using tunable lasers and pointwise full-field 3D scanning of a macroscopic object with an optically rough surface (plaster bust). However, \cite{dandliker1988two, li2018sh} still rely on raster scanning to capture a full-field 3D image, which limits acquisition speed and may introduce alignment errors. 

 From the above discussion, it becomes clear that current state-of-the-art ToF based approaches are poorly suited for measuring real world objects (e.g., human faces, plaster statues, and technical parts) at high speeds (i.e. without relying on raster scanning) and with high depth precision (i.e. $\delta z < 1mm$). In this work, we propose a novel 3D sensor that fills the gap between single-wavelength interferometry and ToF cameras and enables 3D measurements of real world objects with high resolution, high data density (point cloud density), and fast data acquisition (see Fig.~\ref{fig:scope}(a)). Our approach is based on multi-wavelength interferometry, i.e. it exploits wavelength diversity. We demonstrate the proposed approach to interferometric 3D sensing using three distinct focal-plane array (FPA) architectures, including a lock-in sensor, a flutter shutter sensor, and a commercial off-the-shelf CMOS sensor. The proposed embodiments differ in lock-in frequencies, data capture speeds, and point cloud densities as shown in Fig.~\ref{fig:scope}(b).

\begin{figure*}[htpb]
    \centering
    \includegraphics[width = 1\linewidth]{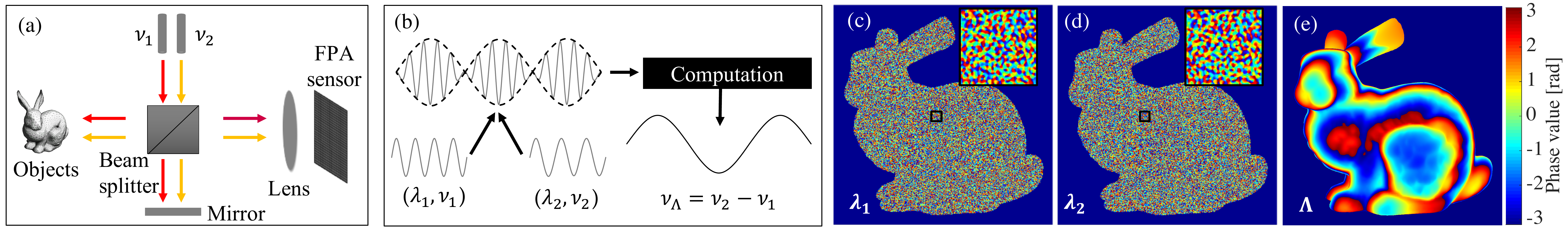}
    \caption{(a) Our proposed FPA-based ToF 3D sensor leverages optical interferometry with wavelength diversity. (b) Principle of wavelength diversity: the laser emissions are electromagnetic waves with frequencies of $\nu_1$ and $\nu_2$. After mixing these two waves, there are a high frequency component ($\nu_1$+$\nu_2$) and a low frequency one (|$\nu_1$ - $\nu_2$|). The proposed 3D sensor computationally selects out the frequency of $|\nu_1 - \nu_2|$ as the measurement frequency providing a synthetic wavelength $\Lambda$ from several meters to several micrometers. The phase map of a rough surface bunny is random due to speckles with single optical wavelengths $\lambda_1$ (c) and $\lambda_2$ (d). However, we can recover the phase (depth) of the same bunny with a synthetic wavelength $\Lambda$ (e) with interferometric depth resolution using the proposed 3D sensor. These are simulated results.}
    \label{fig:highLevel}
\end{figure*}


\subsection{Basic principle of the proposed method}
Our FPA-based ToF 3D sensor (Fig.~\ref{fig:highLevel}(a)) leverages ideas from multi-wavelength interferometry to provide high temporal, lateral, and depth resolution for real world objects. Two optical wavelengths ($\lambda_1$ and $\lambda_2$) are used simultaneously in the system. By computationally mixing (interfering) the two captured optical fields (frequencies $\nu_1$ and $\nu_2$) and filtering the  high frequency component ($\nu_1+\nu_2$), we are able to interrogate the scene at the relatively low beat-note frequency $\nu_{\Lambda} =|\nu_1-\nu_2|$, which defines a "Synthetic Wavelength" $\Lambda$ as
\begin{equation}
    \Lambda = \frac{c}{\nu_{\Lambda} } = \frac{c}{|\nu_1-\nu_2|}
\label{eq:SWLFreq}
\end{equation}
where $c$ is the speed of light. The synthetic frequency $\nu_{\Lambda}$ serves as the measurement frequency in our proposed method. To isolate this frequency, we pair heterodyne and superheterodyne detection methods with computational algorithms (see more details in Sec. 3~\ref{LockinPrinciple}).

Measuring optically rough surfaces at the synthetic wavelength ${\Lambda}$ yields distinct benefits compared to interferometric measurements with a single optical wavelength: Since the surface roughness is commonly much larger than the optical wavelength $\lambda$, the phase map (which encodes the depth) measured with single-wavelength interferometry approaches is randomized and corrupted by speckle~\cite{goodman1975statistical} (see Fig.~\ref{fig:highLevel}(c) and (d)). Since, however,  the synthetic wavelength can be chosen much larger than the surface roughness, the "synthetic phase map" is robust against speckle~\cite{ willomitzer2019high, willomitzer2019synthetic}, which allows us to recover the related phase information at the synthetic wavelength, exploiting the principles shown in Fig.~\ref{fig:highLevel}(e). Due to the tunability of $\lambda_1$ and $\lambda_2$ in our lasers, the synthetic wavelength can be freely chosen in a range spanning from several meters to several microns. As the synthetic wavelength $\Lambda$ is directly proportional to the unique measurement range of the ToF sensor and inversely proportional to its depth precision \cite{li2018sh,willomitzer2019synthetic}, several realizations of ToF sensors applied to different scenarios are possible. Moreover, the precision of the measurement can be increased by sequentially measuring the scene at two different synthetic wavelengths, followed by multi-frequency phase unwrapping of the measured synthetic phase maps~\cite{willomitzer2019synthetic}.

As previously mentioned, our proposed sensor concept does not require raster scanning as is the case in the previous work ~\cite{li2018sh}. Moreover, it measures the object surface in a deterministic fashion without relying on prior knowledge as in the previous work~\cite{wished2020}, exploiting only a fraction of the necessary sequentially acquired images.

\subsection{Contributions}
The unique contributions of our work are summarized below:

\begin{itemize}
\item \textbf{Focal-plane array based ToF 3D sensor with megapixel chip resolution}: To the best of our knowledge, the experiments disclosed in this contribution represent the first documented demonstration of focal-plane based heterodyne/superheterodyne interferometry for 3D imaging with megapixel chip resolution.
\item
\textbf{ToF 3D sensing with tunable depth resolution in the micron-millimeter range:} We demonstrate depth resolutions that far exceed CAMW ToF sensors, yet are also significantly more robust to speckle than purely interferometric techniques. 
\item
\textbf{Three different ToF camera prototypes exploiting different pixel architectures:} We demonstrate prototype implementations using three full-field depth sensor concepts that differ in the choice of focal plane architecture, specifically detection complexity and pixel resolution. Performances with these focal-plane sensors are evaluated and compared, and the benefits and drawbacks of each implementation are discussed.
\end{itemize}


\section{Experimental setup}
\label{sec:setup}
\begin{figure*}[ht!]
    \centering
    \includegraphics[width = 1\linewidth]{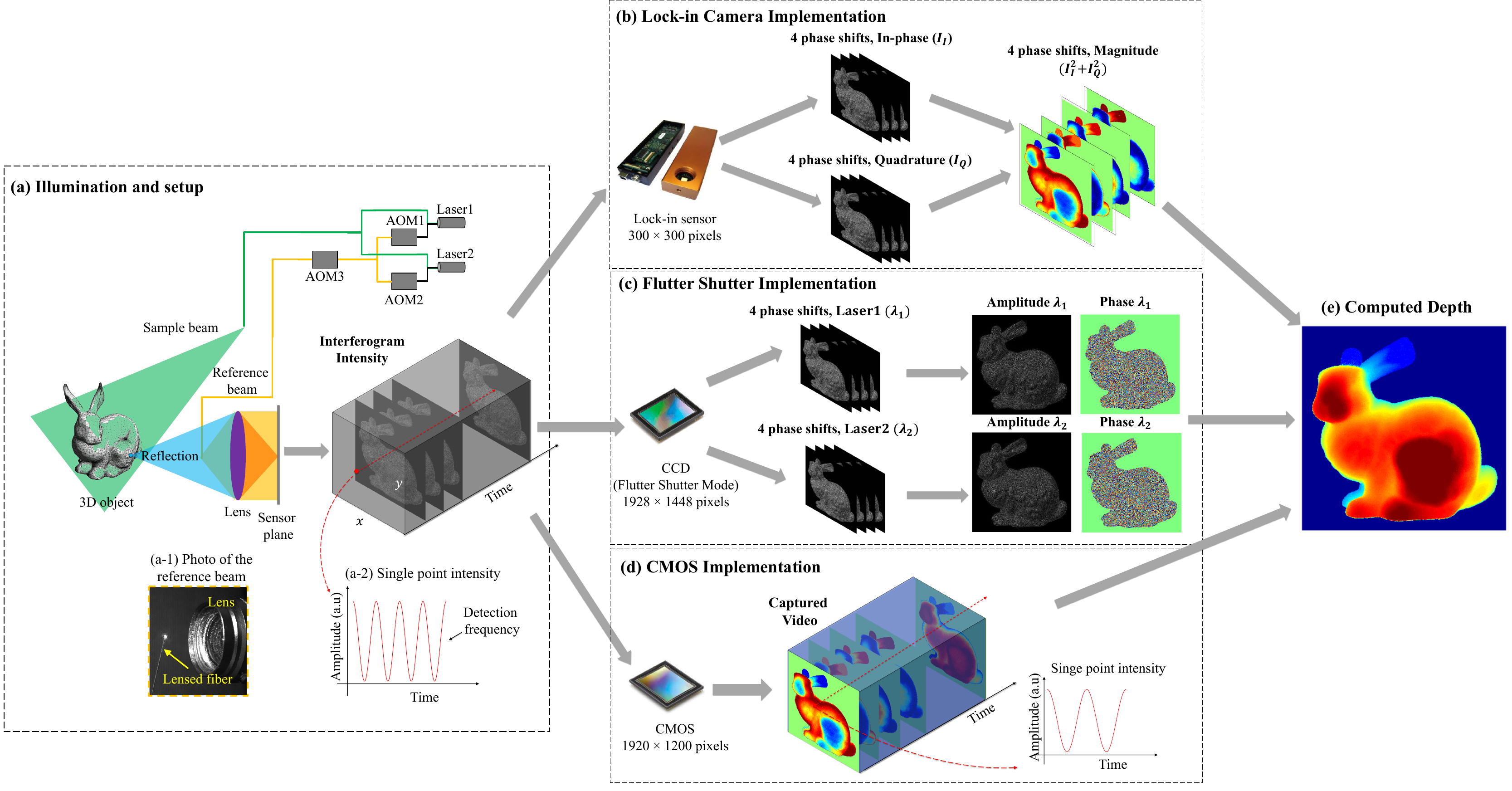}
    \caption{\textbf{Implementation of the proposed 3D sensor with different FPA imagers.} (a) Experimental setup: two lasers with closely spaced wavelengths are split into reference (brown) and sample (green) beams. In the reference beam, we use three AOMs to introduce a fixed frequency shift as the detection frequency (more details in Sec.~\ref{sec:setup}). The reference beam is directed via a custom fiber to the front focal point of the lens (shown in the photo (a-1)). The reflection interferes with the reference beam in front of the sensor plane, producing an interferogram that varies with the detection frequency (a-2). (b) \textbf{Prototype with a lock-in imager:} The sensor demodulates the interferogram and outputs in-phase ($I_I$) and quadrature ($I_Q$) terms. We get $I_I^2+I_Q^2$ by squaring and summing these two outputs. We then apply a four-phase-shifting method to compute the depth with synthetic wavelength. (c) \textbf{Prototype with a flutter shutter imager (high lateral resolution):} The phase is computed sequentially for each wavelength, and depth is computed from the difference in phase between the two wavelengths. (d) \textbf{Prototype with a CMOS imager (high lateral resolution and low lock-in frequency):} The imager samples the interferogram field at the lock-in frequency in the time domain. We computationally square the sensor output and pass through a temporal band-pass filter. This produces a temporal varying sinusoidal signal that allows depth to be computed from the synthetic wavelength as shown in (e). The results of bunny are simulated.}
    \label{fig:sensorCompare}
\end{figure*}

A schematic of the proposed 3D sensor setup is shown in Fig.~\ref{fig:sensorCompare}(a). Two tunable lasers (Toptica DFB pro 855nm) with closely spaced wavelengths ($\lambda_1$, $\lambda_2$) are turned on simultaneously, and each laser emission is fed into a fiber beam splitter (Thorlabs PN850R3A2) to produce sample (green) and reference (brown) beams. Sample beams from two lasers are combined with a fiber beam coupler (Thorlabs PN850R5A2) and illuminate the object from the same point (fiber tip). In the reference arm, we use acousto-optic modulators (AOM1  and AOM2, IntraAction Corp) to shift up the frequency of the electromagnetic fields propagating in the fiber by $f_1$. A phase delay $\Delta \phi$ is added between the driving signals of AOM1 and AOM2, leading to a fixed phase offset $\Delta \phi$ in the reference beam of laser 1 with respect to laser 2~\cite{malacara2007optical,li2018sh,willomitzer2019synthetic}. Eventually, reference beams from the two lasers are combined with a fiber coupler and directed through AOM3 to downshift the frequency by $f_3$. As a result, a fixed frequency shift of $f = f_1-f_3$  is established between the reference and sample beams for each laser. Lastly, the combined reference beam is directed onto the sensor using a customized lensed fiber needle (WT\&T Inc.) positioned in the front focal point of the objective lens ~\cite{willomitzer2019synthetic} as shown in Fig.~\ref{fig:sensorCompare}(a). The lensed fiber avoids using a beam splitter to combine reference and sample beams, which increases the light throughput and makes the system more compact.

The reflection from the object is imaged onto the sensor where it interferes with the reference beam. We can derive an expression for the instantaneous irradiance (Fig.~\ref{fig:sensorCompare}(b)) at point (x,y) on the sensor plane as below.
\begin{equation}
\small
\begin{split}
I(x,y,t) &=a_0(x,y)+a_1(x,y)\cos{\left( \phi_{\lambda_1}(x,y)-2\pi ft+ \Delta \phi \right)}\\
&+a_2(x,y)\cos{\left( \phi_{\lambda_2}(x,y)-2\pi ft \right )}
\end{split}
\label{Eq_I_all}
\end{equation}
where $a_0$ is scalar which represents the sum of intensity in sample and reference beams of laser 1 and 2. $a_1$ and $a_2$ are scalars representing the amplitude of the interferogram for laser 1 and 2 respectively. $f = f_1-f_3$ is the AOM beat-note frequency and $\Delta \phi$ is the phase offset introduced between  AOM1 and AOM, later used for phase shifting of the synthetic interferogram~\cite{schwider1989phase, malacara2007optical}. $\phi_{\lambda_1}(x,y)$ and $\phi_{\lambda_2}(x,y)$ are the captured phasemaps at the optical wavelength $\lambda_1$ and $\lambda_2$. Note that these phasemaps are stochastic in nature (a.k.a speckles) and would individually not reveal any information about the optical path difference (OPD) between the sample and reference arms. This is the case because the optical path in the sample arm encapsulates the combined effect of macroscopic depth variations associated with topographic changes in the object and microscopic height variations due to surface roughness of the object. \textit{Only} the phase variations associated with the macroscopic surface variations are of interest from the standpoint of macroscopic 3D imaging. With our principle, we are able to isolate these phase variations from the random phase variations due to speckle. This is realized by calculating a "synthetic phase map" $\phi_{\Lambda}(x,y)$ from the randomized (but correlated!) phase maps $\phi_{\lambda_1}(x,y)$ and $\phi_{\lambda_2}(x,y)$:
\begin{equation}
\begin{split}
\small
\phi_{\Lambda}(x,y) &= h\left (\phi_{\lambda_1}(x,y),\phi_{\lambda_2}(x,y) \right)\\
&= \frac{4\pi L(x,y)}{\Lambda} = \frac{4\pi L(x,y)}{c}\cdot \nu_{\Lambda}
\label{eq:SyntPhaseMap}
\end{split}
\end{equation}
\noindent
$h(.)$ is a function to estimate the synthetic phase map from interferograms at the two wavelengths ($\lambda_1, \lambda_2$).  $L$ is the optical path difference (OPD) between the sample and reference arms, that can now be reconstructed since $\phi_{\Lambda}(x,y)$ is not subject to speckle anymore (see Fig.~\ref{fig:highLevel}(e)). 
As it follows from Eq.~\ref{eq:SyntPhaseMap}, our proposed imaging process is very similar to the process in a ToF camera, however at a much higher modulation frequency $\nu_\Lambda$ (GHz to THz instead of MHz in conventional ToF cameras), which leads to higher depth precision. We will additionally show in the next sections that the detection frequency of our sensors (lock-in frequency) does not have to match $\nu_\Lambda$ and can be picked much lower. This is another important point that distinguishes our approach from conventional ToF cameras and allows us to use (slow) off-the-shelf CCD/CMOS cameras for high precision ToF-based 3D measurements (see Sec.~\ref{implementation_cmos}).

In subsequent sections, we examine three depth sensor prototypes using FPA imagers with different pixel architectures and pixel sizes including a lock-in sensor (HeliCam C3), a CCD camera (FL3-GE-28S4M-C, flutter shutter mode), and a CMOS camera (GS3-U3-23S5M-C). For each sensor architecture, we specify the function $h(.)$, i.e. we describe how to calculate the synthetic phase map from the detected optical signals. Since the calculation of the synthetic phase map is a pixel-wise process, we drop the reference to pixel coordinates (x,y) for the rest of the paper.

\section{Implementation with a lock-in sensor}
\label{ImplementWithLS}
The first prototype described herein uses a lock-in FPA camera~\cite{lockInSensor} wherein each pixel features the ability to electronically demodulate the received irradiance with an RF frequency $f$. The lock-in sensor used in this paper has a sensor resolution of 300$\times$300 pixels. The exposure time of each measurement is 23ms, corresponding to 70 cycles of the RF frequency $f$ = 3KHz. After detailing the detection scheme, we demonstrate the prototype on different objects, including translucent and opaque objects which are challenging to be imaged with single-wavelength optical interferometry.

\subsection{Principle}
\label{LockinPrinciple}
Eq.~\ref{Eq_I_all} states that the irradiance at the sensor plane is a superposition of two cosines temporally varying at the frequency $f$. To detect the signal, we set the lock-in sensor demodulation frequency to $f$. For smooth object surfaces, the time-independent phase shift $\phi_{\lambda_{1,2}}$ associated with each cosine encodes the depth information $L$ (modulo $\lambda_{1,2}$) via $\phi_{\lambda_{1,2}} = \frac{4\pi L}{\lambda_{1,2}}$. As discussed, $\phi_{\lambda_{1,2}}$ is subject to speckle for rough object surfaces, which makes it impossible to retrieve $L$ from a single-wavelength measurement.  

Each lock-in sensor pixel produces two outputs: one "in-phase" term ($I_I$) and one  "quadrature" term ($I_Q$)~\cite{willomitzer2019synthetic, cossairt2019micro}: 

\begin{equation}
\small
I_I =m\cos \left(\phi_{\lambda_1}+ \Delta\phi\right) + n\cos \left(\phi_{\lambda_2} \right)
\label{lock_in_in_phase_eq}
\end{equation}

\begin{equation}
\small
I_Q = m\sin \left(\phi_{\lambda_1}+ \Delta\phi\right) + n\sin \left(\phi_{\lambda_2} \right)
\label{Eq_I}
\end{equation}

\noindent where $m$ and $n$ are scalars.

We computationally combine in-phase and quadrature-phase readouts to estimate the depth at each pixel using the synthetic wavelength ($\Lambda = c/|\nu_1-\nu_2|=\lambda_1 \cdot \lambda_2/|\lambda_1-\lambda_2|$). 

\begin{equation}
\small
I_{rc}(\Delta \phi) = I_I^2 + I_Q^2
= 2mn\cos\left(\mathop{\underbrace{\phi_{\lambda_1}-\phi_{\lambda_2}}}_{\phi_{\Lambda}}+ \Delta\phi \right)+\left(m^2+n^2\right)
\label{lock_in_combine_eq}
\end{equation}

Eq.~\ref{lock_in_combine_eq} can be viewed as an \textit{interferogram at the synthetic wavelength}, where $\phi_{\Lambda}$ encodes the depth $L$ via  $\phi_{\Lambda}= \frac{4\pi L}{\Lambda}$. $\Delta \phi$ is the absolute phase shift between $\lambda_1$ and $\lambda_2$, which can be freely changed. After each acquired measurement, we change $\Delta \phi$ by $90^\circ$ and subsequently use the four-phaseshift algorithm to compute the depth $L$ via

\begin{equation}
L = \mathrm{atan}\left(\frac{I_{rc}(270^{\circ})-I_{rc}(90^{\circ})}{I_{rc}(0^{\circ})-I_{rc}(180^{\circ})}\right)\cdot \frac{1}{4\pi}\cdot \Lambda
\label{eq:depth}
\end{equation}

Additionally, the albedo $M$ of the object  can be also recovered:
\begin{equation}
M = \sqrt{(I_{rc}(270^{\circ})+I_{rc}(90^{\circ}))^2+(I_{rc}(0^{\circ})+I_{rc}(180^{\circ}))^2}
\label{eq:intensity}
\end{equation}

This process happens in parallel in each pixel of the lock-in camera. It becomes obvious that, with respect to  previously published work~\cite{dandliker1988two,li2018sh}, the availability of a lock-in FPA sensor eliminates the need for raster scanning which increases the temporal bandwidth of the 3D sensor.

\subsection{Imaging translucent objects}
\begin{figure}[htp!]
    \centering
    \includegraphics[width = 1\linewidth]{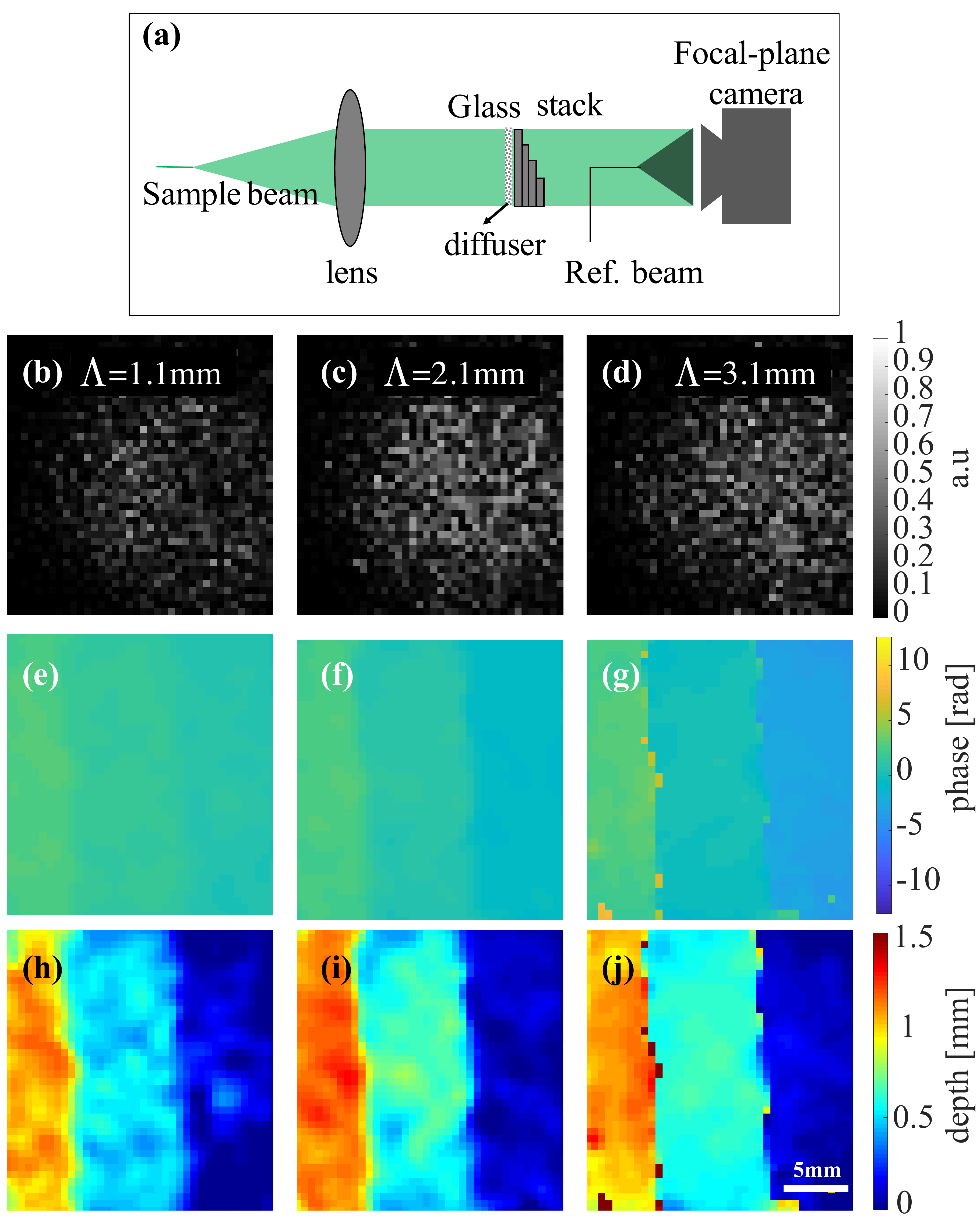}
    \caption{Imaging a glass stack with a diffuser. (a) shows the test setup for the translucent object. (b-d) show intensities captured by the lock-in sensor under three different synthetic wavelengths. (e-g) are estimated phase maps with corresponding synthetic wavelengths of 1.1mm, 2.1mm, and 3.1mm. (h-j) shows OPD/depth information.}
    \label{fig:exp_tran_glass}
\end{figure}

A stack of glass plates with a diffuser in front (Thorlabs glass diffuser 220 Grit, see Fig.~\ref{fig:exp_tran_glass}(a)) is measured with the first prototype. The sample beam from the fiber is collimated with a lens before illuminating the object. The translucent object is measured at three different synthetic wavelengths: 1.1mm, 2.1mm, and 3.1mm. The amplitude images taken with the lock-in sensor are shown in Fig.~\ref{fig:exp_tran_glass}(b-d). The diffuser randomizes the wavefront of the illumination beam and introduces subjective speckle  in the intensity/amplitude images. We note that \textit{the subjective speckle pattern is not resolved by the imager and a finite number of subjective speckles is averaged over one pixel}, as  it can be seen in Fig.~\ref{fig:exp_tran_glass}(b). This imaging configuration is chosen deliberately. The related theoretical considerations will be discussed in Sec. ~\ref{discussion1}). 

The physical thickness of each glass plate is 1mm and the refractive index of the glass is about 1.5. Therefore, each glass plate introduces an OPD of 0.5mm. It can be seen in the calculated synthetic phase maps Figs.~\ref{fig:exp_tran_glass}(e-g) that a large synthetic wavelength leads to a smaller phase variation while the variations in phase are more prominent for shorter synthetic wavelengths. Consequently, a shorter synthetic wavelength provides better depth resolution compared to the utilization of a large synthetic wavelength. This statement will be quantitatively confirmed in the next experiment. However, the clear separability of the $0.5mm$ OPD introduced by each glass stack already alludes to the sub-mm depth precision capabilities of our ToF approach.  

\subsection{Imaging opaque objects}
In a second experiment, we measure the surface of a planar cardboard (Fig.~\ref{fig:sensorCompare}(a)) to further evaluate the depth precision of our approach. The cardboard surface is optically rough and a fully developed subjective speckle pattern is visible in the amplitude image of the lock-in camera. We test with four different synthetic wavelengths: 69.30mm, 33.54mm, 8.69mm, and 1.08mm. 

\begin{figure}[htp!]
    \centering
    \includegraphics[width = 1\linewidth]{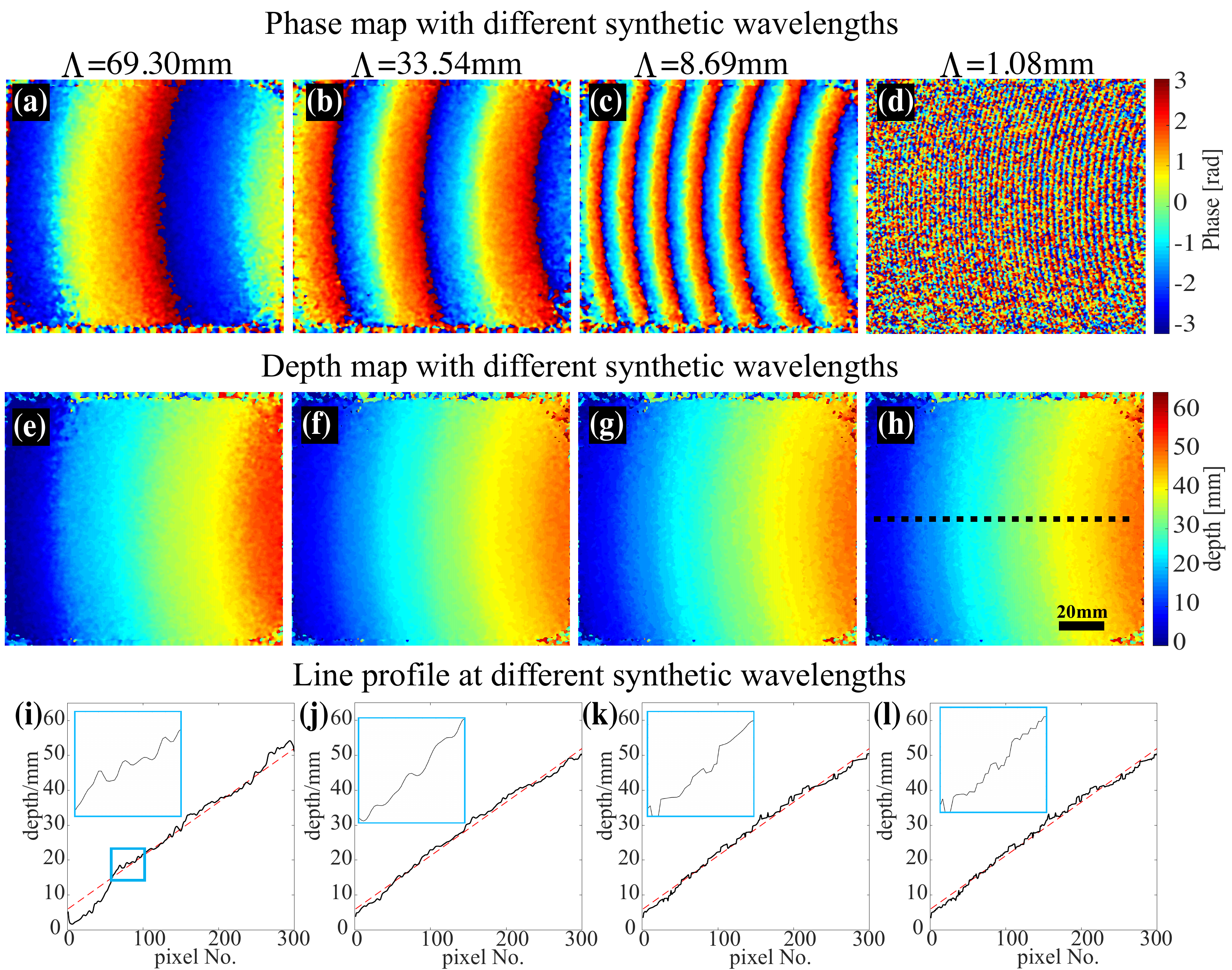}
    \caption{Imaging a planar cardboard using the prototype with the lock-in sensor: (a-d) show phase maps at different synthetic wavelengths. (e-h) show ray depth maps after phase unwrapping process. (i-l) are the line-profiles of the ray depth maps across the cardboard. We fit a line curve (red) to the measurements, and quantify the RMSE values between measurements and fitted curve as 2.384mm, 1.165mm, 1.192mm, and 1.222mm for synthetic wavelengths of 69.30mm, 33.54mm, 8.69mm, and 1.08mm, respectively.}
    \label{fig:exp_cardboard}
\end{figure}

The fiber tip illuminates the cardboard at an angle, resulting in an off-axis Fresnel zone plate for the synthetic phase map (Figs.~\ref{fig:exp_cardboard}(a-d)). To quantify the depth precision of the present prototype and  the proportionality of synthetic wavelength and depth precision we repeatedly measured the same point on the cardboard for 50 times at different synthetic wavelengths, respectively. Eventually, we calculated the standard deviation over the 50 measurements for each wavelength. The results are shown in Tab.~\ref{table:cardboard}. While all measurements display roughly the same phase error $\delta \phi$, the trend to higher precision for smaller synthetic wavelengths is clearly visible. For the smallest used synthetic wavelength of $\Lambda = 1.08mm$, our prototype achieves a depth precision of only $\delta z = 35 \mu m$.

\begin{table}[htp!]
\centering
\small
{
\begin{tabularx}{\linewidth}{m|s|s|s|s}\hline
$\Lambda$ [mm]&69.3&33.54&8.69&1.08\\\hline
$\delta \phi$ [rad] & 0.1155 & 0.0844 &0.1030 & 0.2070 \\ 
\hline
$\delta z$ [mm] & 1.274 & 0.450 & 0.142 & 0.035 \\  
\hline
\end{tabularx}
}
\caption{We show standard deviations of phase ($\delta \phi$) and ray depth ($\delta z$) of a fixed point on the object for 50 repeated measurements with corresponding synthetic wavelengths ($\Lambda$). This process is performed on the wrapped phase estimations in Fig.~\ref{fig:exp_cardboard}(a-d).}
\label{table:cardboard}
\end{table}

After considering only a single point on the measured object (a single pixel in our $300 \times 300 pix$ camera image) for 50 independent measurements, we investigate the full-field capture of the entire object surface for one single measurement, however, still at the same synthetic wavelengths. The process described above also delivers synthetic phase maps (Figs.~\ref{fig:exp_cardboard}(a-d)), one for each measured synthetic wavelength. The phase maps are wrapped due to the geometrical arrangement of the experiment.

 We use continuity constraints to  manually unwrap the phase map for $\Lambda = 69.3mm$ (Fig.~\ref{fig:exp_cardboard}(a)) and use the received unwrapped phase map  as a guidance to unwrap the phase maps acquired at smaller synthetic wavelengths (Fig.~\ref{fig:exp_cardboard}(b-d)), using the multi-frequency unwrapping process described in~\cite{xu1994phase, gupta2015phasor}. Eventually, we convert the unwrapped phase maps into depths as shown in Fig.~\ref{fig:exp_cardboard}(e-h). A  profile (see Fig.~\ref{fig:exp_cardboard}h) of the measured depth  maps  is plotted as black line in Fig.~\ref{fig:exp_cardboard}(i-l). We calculate the precision for each profile by subtracting a best fit plane from the data and calculate the standard deviation. It is expected that the measurements at smaller synthetic wavelengths will exhibit smoother line profiles. However, this is not the case: The evaluated depth precision values are  2.384mm, 1.165mm, 1.192mm, and 1.222mm for synthetic wavelengths of 69.30mm, 33.54mm, 8.69mm, and 1.08mm, respectively, meaning that the precision of the profile is not further improved for synthetic wavelengths smaller than 8.69mm and 1.08mm. This effect might be caused by noise propagation during phase unwrapping and is subject to further investigation.

\begin{figure*}[htp!]
    \centering
    \includegraphics[width = 1\linewidth]{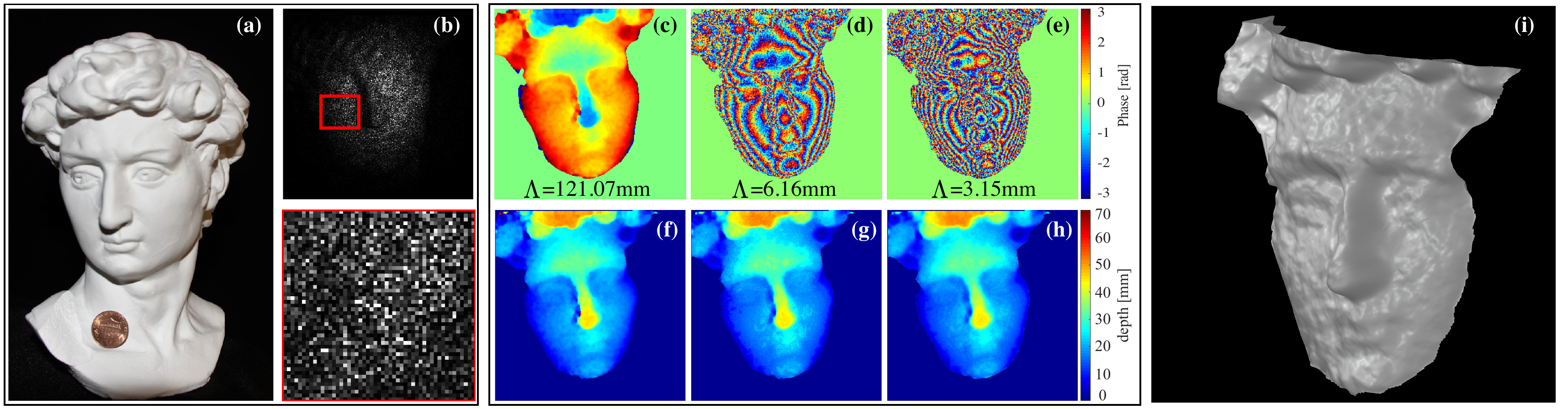}
    \caption{Imaging a plaster bust with the prototype of a lock-in sensor. (a) Photo of the plaster. (b) Intensity measurement using the lock-in sensor with closeup marked with red box. (c-f) Phase maps with different synthetic wavelengths of 121.07mm, 6.16mm, and 3.15mm. (f-h) Depth maps after unwrapping process with corresponding synthetic wavelengths. (i) Rendered depth based on the depth map of (h).}
    \label{fig:david_amp}
\end{figure*}

In a last experiment for this sensor configuration, we scan a three-dimensional plaster bust (Fig.~\ref{fig:david_amp}(a), dimensions $\sim$ 11mm $\times$ 16mm $\times$ 12mm ). To eliminate the angular carrier in  the acquired synthetic phase map the illumination and observation are co-located for this experiment.   Fig.~\ref{fig:david_amp}(b) shows the measured amplitude image (speckle is not resolved). Synthetic phase maps at different synthetic wavelengths of 121.07mm, 6.16mm, and 3.15mm are captured (Fig.~\ref{fig:david_amp} (c-e)).

The phase map at the largest synthetic wavelength (Fig.~\ref{fig:david_amp}(c)) is used as a guidance to unwrap the phase maps at higher frequency (Fig.~\ref{fig:david_amp}(d and e)). Eventually, the depth map is calculated from the unwrapped phase map Fig.~\ref{fig:david_amp}(f-h). Albeit the high depth precision, the pixel resolution of the lock-in sensor (only $300 \times 300$ pixels), and consequently the lateral resolution of the acquired 3D model leaves room for improvement. For the shown result, this prevents us from seeing high frequency details around mouth and eyes of the bust. 

\section{Implementation with a flutter shutter camera}
As mentioned before, our novel sensor concept does not require the lock-in frequency $f$ to be matched with the synthetic frequency $\nu_{\Lambda} = \nu_1-\nu_2$ that defines the depth precision of the measurement. We note again that this match necessary in conventional ToF cameras, where it represents the \textit{ limiting factor for depth precision}. The described relaxation removes the requirement for high detection frequencies and opens doors to use regular CCD/CMOS sensors with much higher lateral resolutions. This will be discussed in the following.

In a second embodiment, we employ a "flutter shutter sensor"~\cite{raskar2006coded,balaji2017spatiotemporal} which is a special form of a global shutter CCD camera. During one integration cycle, the global shutter is electronically chopped multiple times, allowing for a temporal modulation of the exposure. 

\subsection{Principle}
As described in ~\cite{raskar2006coded}, we model the flutter shutter signal $\sqcap (t)$ as:  $\sqcap (t)$ = $1/2 \cdot (1+ \cos(2\pi ft)+ 1/3 \cdot \cos(2\pi 3f t)+1/(2k-1) \cdot \cos(2\pi(2k-1) ft))$ where $f$ is the demodulation frequency and $k$ is an integer. The demodulation frequency can be low ( several KHz). The output of the flutter shutter camera is then described as:

\begin{equation}
\small
I_f=\int_{0}^{\Delta t}  I(t)\cdot \sqcap (t) dt = m\cos \left(\phi_{\lambda_1}+\Delta \phi \right) + n\cos \left(\phi_{\lambda_2} \right) + D\cdot \Delta t
\label{Eq:FScamera}
\end{equation}
where $I(t)$ is the irradiance at the detector, $\Delta t$ is the exposure time, and $D\cdot \Delta t$ is constant due to the integral of DC components in the interferogram. The integration of high frequency components ($2(k-1)f$) goes to zero after the exposure.

The sensor readout resembles the in-phase readout of the lock-in sensor (Eq.~\ref{lock_in_in_phase_eq}) with the exception of the D.C component $D\cdot \Delta t$. In the following, we propose a solution that enables the flutter shutter camera to behave like the lock-in sensor: We store three demodulation signals of $\sqcap(t)$, $\sqcap(t+\pi/2)$ and $\sqcap(t+\pi)$ on the camera memory. Using these three demodulation signals, we receive three measurements of $I_1 = m\cos \left(\phi_{\lambda_1}+\Delta \phi \right) + n\cos \left(\phi_{\lambda_2} \right) + D\cdot \Delta t$, $I_2 = -m\sin \left(\phi_{\lambda_1}+\Delta \phi \right) - n\sin \left(\phi_{\lambda_2} \right) + D\cdot \Delta t$, $I_3 = -m\cos \left(\phi_{\lambda_1}+\Delta \phi \right) - n\cos \left(\phi_{\lambda_2} \right) + D\cdot \Delta t$. Using $I_1$ and $I_3$, we can estimate the ambient component $D\cdot \Delta t$ and subtract the DC component from $I_1$ and $I_2$. The result is the same output (Eq.~\ref{lock_in_combine_eq}) as received for the sensor embodiment using a lock-in sensor. To calculate the depth with the four phase-shifting method, a total of 9 measurements is required. An alternative that requires less measurements but more hardware is to split the observation with a beam splitter and use two flutter shutter cameras that are synchronized by a fixed phase offset. 

We implement the prototype with a single flutter shutter camera and illuminating the object with one wavelength on at a time. In subsequent discussions, we restrict attention to the use of a single flutter shutter sensor and sequential acquisition of the interferogram. The sensor readout at each pixel can be written as:

\begin{equation}
\small
I_{\lambda_{k}}(\Delta \phi) =\int_{0}^{\Delta t}  I(t)\cdot \sqcap (t) dt = m\cos \left(\phi_{\lambda_{k}}+ \Delta \phi \right) + D\cdot \Delta t, k= 1,2
\label{Eq_I}
\end{equation}

With this configuration, we first estimate the fields $E_{\lambda_k}$ for each optical  wavelength $\lambda_{k}$
\begin{equation}
\small
E_{\lambda_k} = \frac{I_{\lambda_k}(0^{\circ})-I_{\lambda_k}(180^{\circ})}{2} + i\cdot \frac{I_{\lambda_k}(270^{\circ})-I_{\lambda_k}(90^{\circ})}{2}, k = 1,2
\label{Eq:FS_hologram}
\end{equation}

\noindent and eventually correlate the optical field measurements to obtain the synthetic phase map and depth map: 

\begin{equation}
\small
L = \angle \left (E_{\lambda_1}\odot {E_{\lambda_2}^*}  \right) \cdot \frac{1}{4\pi}\cdot \Lambda
\label{Eq:FS_depth}
\end{equation}

\subsection{Imaging opaque objects}
The flutter shutter camera used in this experiment has 1448$\times$1928 pixels. The demodulation frequency of 5KHz is provided by an external function generator, and each shutter time is 0.1ms. We acquire 60 shutter cycles for each exposure. 

\begin{figure}[h!]
    \centering
    \includegraphics[width = 1\linewidth]{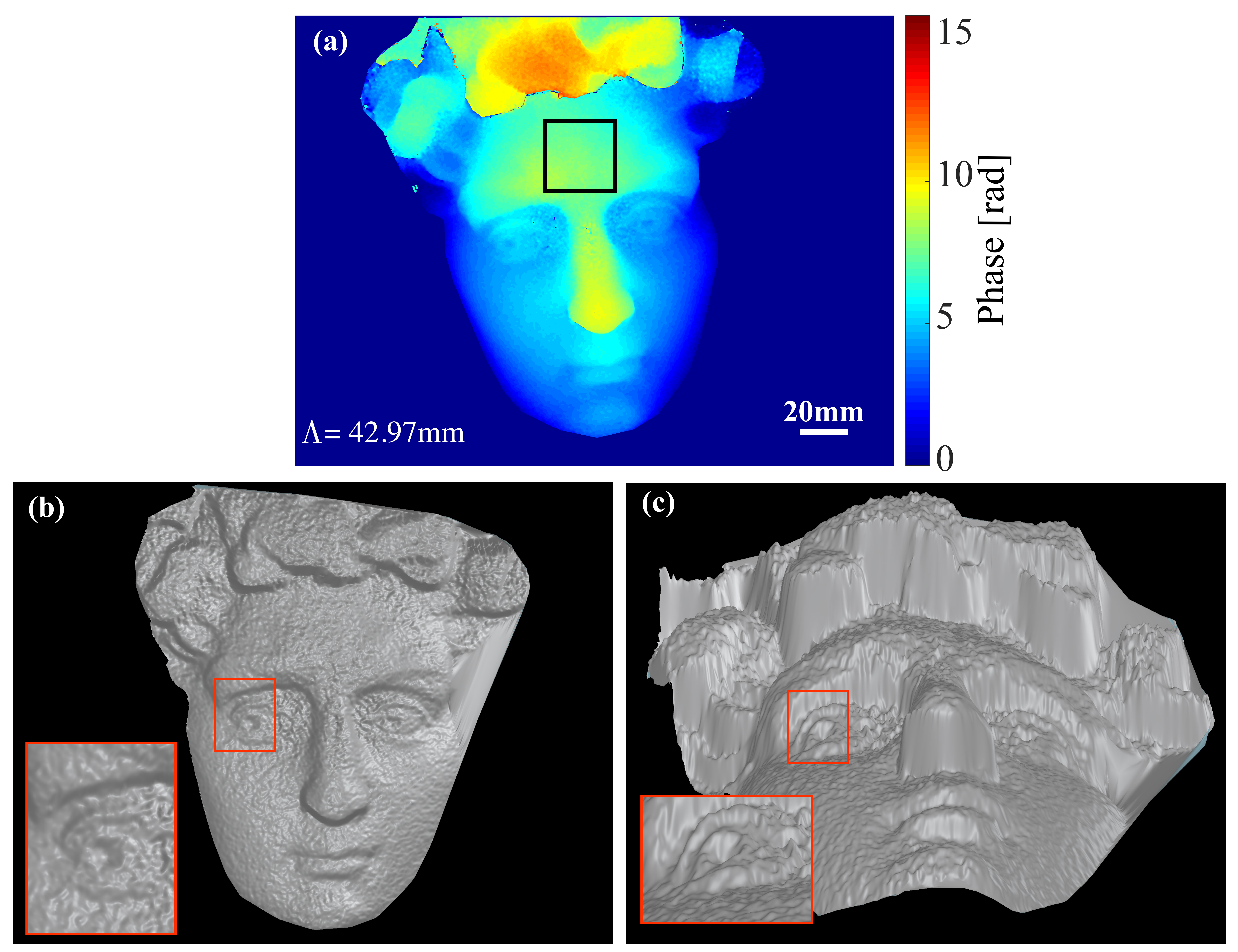}
    \caption{Imaging plaster bust with the prototype of the flutter shutter camera: (a) shows the phase maps with $\Lambda$ of 42.97mm. We render the scan from different view angles: front view (b) and side view (c). We choose a relatively flat region (200$\times$200 pixels) marked with black box in (a) and quantify its standard deviation as 0.376mm.}
    \label{fig:exp_FS_david}
\end{figure}

Utilizing this setup, we measure the plaster bust used in previous experiments at a synthetic wavelength of 43mm. The captured synthetic phase map is shown  in Fig.~\ref{fig:exp_FS_david}(a). The respective depth map is rendered in  Figs.~\ref{fig:exp_FS_david}(b,c) from two viewing angles. Compared to the 3D results shown in Fig.~\ref{fig:david_amp}, the improvement in lateral point cloud resolution is striking! Fine details such as the eye and mouth of the bust can be nicely resolved. The depth precision of the measurement is estimated by calculating the standard deviation of a smooth surface patch on the forehead of the bust (marked with a black box in Fig.~\ref{fig:exp_FS_david}(a)), after subtracting a low pass filtered version of this  surface patch. The value of 0.376mm is in agreement with our previous measurements and shows that the precision of the principle is independent of the RF detection frequency by utilizing CCD/CMOS cameras with small modulation frequencies.

\section{Implementation with a CMOS camera}
\label{implementation_cmos}
Although the flutter shutter sensor can provide higher lateral resolution (>1 megapixels) in comparison to the lock-in camera, it still requires a custom pixel readout architecture that provides temporal modulation at a relatively high frequency. In this section, we demonstrate how to implement our ToF architecture on standard CMOS sensors. This allows us to create a \textit{novel ToF camera with high depth and lateral resolution exploiting only a commodity focal plane array!} We will see, however,  that this does not come without a tradeoff. Due to the long time (several seconds) that this sensor modification requires to capture one 3D model,  its application is restricted to static objects~\cite{verrier2015full,cervantes2007real, Li:19}.

The image formation process is described in Fig.~\ref{fig:sensorCompare}(d): two lasers are turned on simultaneously and  AOM1, AOM2, and AOM3 are set to be 80MHz+15Hz, 80MHz+23Hz, and 80MHz respectively. This creates an AOM beat note frequency $f_1$=15Hz for the field at $\lambda_1$ (laser 1) and $f_2$=23Hz for the field at $\lambda_2$ (laser 2). The frame rate of the used CMOS camera is  160fps, and a total of 600 frames is captured for each measurement.

The irradiance on the imager plane can be written as

\begin{equation}
\small
I(t) =a_0+a_1\cos{\left(\phi_{\lambda_1}-2\pi f_1t \right)}\\
      +a_2\cos{\left(\phi_{\lambda_2}-2\pi f_2t \right )}
\label{Eq:Eq_I_rgb}
\end{equation}

For each pixel readout on the imager, we calculate $I(t)^2$, yielding to a time-varying signal comprised of multiple frequency components. A band-pass filter is then used to isolate the beat frequency component ($f_{1}-f_{2}$) which encodes  the depth information at the synthetic wavelength $\Lambda$. For the used AOM frequency values, the beat frequency $f_{1}-f_{2}$ is only $8Hz$, which is easily resolvable with an off-the-shelf FPA CMOS camera. We note that different combinations of beat frequencies ($f_1-f_2$) and CMOS frame-rates may be used, depending on application-specific constraints on sensor speed.  After band-pass filtering, the signal can be expressed as
\begin{equation}
\small
B(t)= a\cdot \cos{[2\pi{\mathop{(f_{1}-f_{2})}}t+\frac{4\pi L}{\Lambda}]}
\label{eq2}
\end{equation}
where $a$ is scalar. We compute the phase component of $B(t)$ via Fourier Transform and eventually calculate the depth value with the method described in~\cite{li2018sh}.

\begin{figure}[htp!]
    \centering
    \includegraphics[width = 1\linewidth]{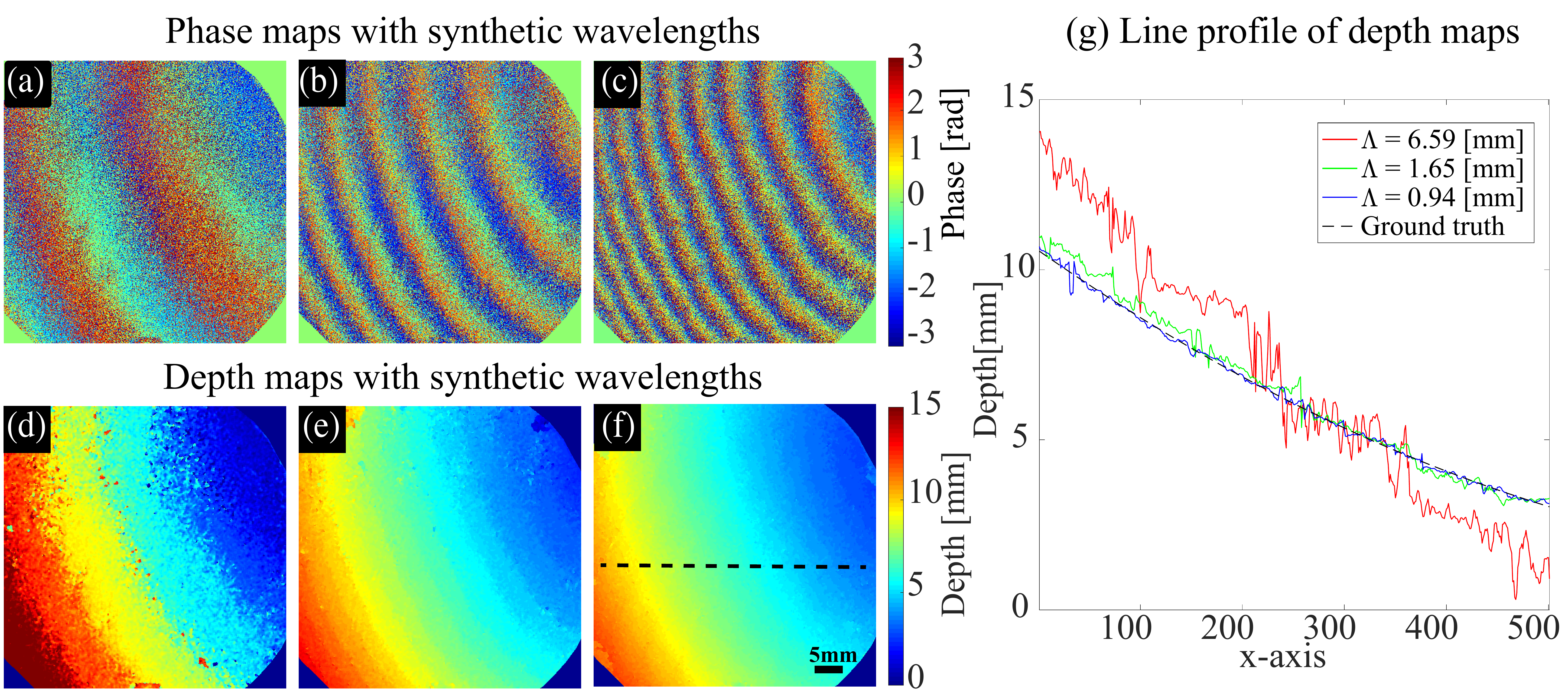}
    \caption{Imaging a piezo disc with the prototype using the CMOS imager. (a-c) Phase maps at synthetic wavelengths $\Lambda$ of 6.59mm, 1.65mm, and 0.94mm. (d-f) Depth maps (after phase unwrapping) at different synthetic wavelengths. (g) Line profiles of depth maps. The RMSE between measurements and fitted ground truth (black dash line) are 1.663mm, 0.292mm, 0.106mm for synthetic wavelengths of 6.59mm, 1.65mm, and 0.94mm, respectively.}
    \label{fig:exp_rgb_plate}
\end{figure}

We utilize the above procedure to measure a metallic planar object (piezo buzzer). The object surface is shiny, but optically rough, and produces a fully developed subjective speckle pattern on the sensor plane. The object surface is measured at three different synthetic wavelengths: 0.94mm, 1.65mm, and 6.59mm. The results are shown in Fig.~\ref{fig:exp_rgb_plate}. Similar to the measurements shown in  Fig.~\ref{fig:exp_cardboard}, a phase ramp is produced due to illumination of the object at an angle. The captured synthetic phase maps (Fig.~\ref{fig:exp_rgb_plate}(a-c)) are unwrapped~\cite{herraez2002fast} and converted into depth maps ( Fig.~\ref{fig:exp_rgb_plate}(d-f)). 

We examine a line trace of the depth profile across the object in Fig.~\ref{fig:exp_rgb_plate}(g) with a fitted curve (marked in dash black line). We quantify the precision of the  measurements as 1.663mm, 0.292mm, and 0.106mm for synthetic wavelengths of 6.59mm, 1.65mm, and 0.94mm, respectively. This preliminary result demonstrates the possibility of implementing a ToF camera with a commodity CMOS sensor. Further experimental validation of this promising approach to 3D sensing should facilitate adjustable and precise power matching to accommodate a variety of objects with varying surface roughness.

\section{Theoretical considerations: Ideal sampling of subjective speckle patterns}
\label{discussion1}
\begin{figure*}[htp!]
    \centering
    \includegraphics[width = 1\linewidth]{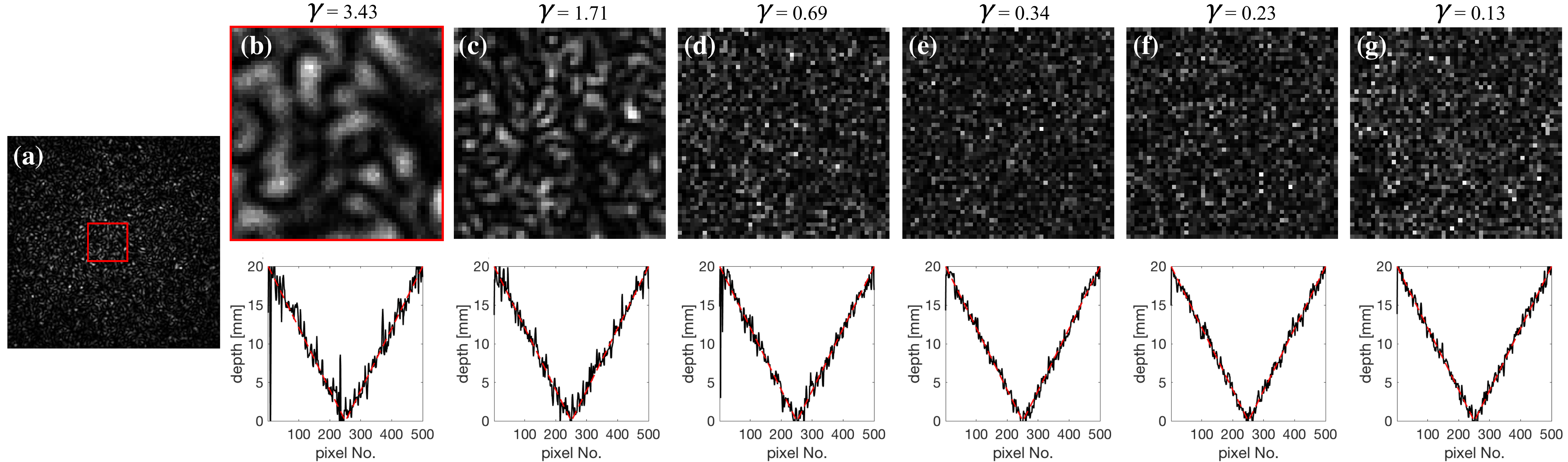}
    \caption{Sampling of speckle versus depth performance for a simulated V-groove. (a) Intensity with a sampling of $\gamma$=3.43 for the full frame and its close up in (b). (c-g) Close-ups of intensities with sampling parameters of $\gamma$=1.71, 0.69, 0.34, 0.23, and 0.13, respectively. Line profiles of depth map with corresponding speckle size are shown below, and red line marks the ground truth.These results are simulated.}
    \label{fig:speckleVSpixel}
\end{figure*}

Our method relies on computationally mixing two electromagnetic fields captured at two slightly different optical wavelengths. Depending on the roughness of the measured surface, these fields may be heavily corrupted by speckle. One of the fascinating characteristics of our method lies in the fact that the random phase variations in a speckle pattern can be canceled out to obtain the phase variations of the synthetic field that are only dependent on the macroscopic depth variations of the measured object surface. The size $\delta_{speckle}$ of the subjective speckles incident on the sensor is dependent on the image-side numerical aperture $NA'$.
\begin{equation}
    \delta_{speckle} = \frac{\lambda}{NA'} 
\label{eq:speckle}
\end{equation}

Most imaging objectives provide the ability to freely adjust the aperture diameter. This poses the question of the \textit{optimal subjective speckle size that leads to the highest depth precision for our introduced approach} arises.
One might think that the subjective speckle fields incident on the detector should be oversampled, i.e. that the size of one subjective speckle should be at least twice the pixel pitch to satisfy Nyquist sampling.

However, this would limit the optical lateral resolution of the method, since the subjective speckle size (Eq.~\ref{eq:speckle}) is also equal to the size of the image-sided diffraction disc. Moreover, a critically sampled subjective speckle field, contains low intensity speckle-minima where the SNR is significantly decreased~\cite{fercher1985rough,li2018sh}. \textit{Can several subjective speckles be averaged over one pixel to improve the SNR?} In the context of our introduced method, this question is not trivial since precise phase values in the captured optical fields are necessary to estimate the synthetic phase map. 

To better understand optimal pixel sampling of subjective speckles in the sensor plane and its effect on the synthetic phase estimation, we perform a simulation with the following key parameters: The simulated FPA sensor has a resolution of 500 $\times$ 500 pixels with a pixel size of 11.2$\mu m$ $\times$ 11.2$\mu m$ ($\delta_{pixel}$). The fill factor is set to 1. We illuminate the scene at two optical wavelengths $\lambda_1= 854nm$ and $\lambda_2= 854.01nm$, resulting in a synthetic wavelength $\Lambda = 72.9mm$. The simulated object consists of two tilted planar surfaces that form a "V-groove". The object has a maximum depth range of 20mm and a surface roughness of $ 8.54 \mu m = 10\lambda_1$. After calculating the subjective speckle fields incident on the sensor plane for each wavelength, we apply white Gaussian noise with an SNR of 30dB to the measurements. We simulate different speckle sizes by varying the $NA'$ of the simulated objective lens. The ratio of speckle size to pixel size is defined by $\gamma=\delta_{speckle}/\delta_{pixel}$.

Fig.~\ref{fig:speckleVSpixel} shows the simulated FPA images and the related retrieved 3D profiles of the V-groove for different values of $\gamma$. A quantitative evaluation of the related depth-RSME is given in Tab. XXX. It can be seen that the RSME decreases for decreasing $\gamma \geq 0.34$. This means that \textit{speckle averaging has a positive effect} on the phase noise of the synthetic phase map until $\gamma$ hits a certain threshold. For $\gamma < 0.34$, the depth-RSME increases with decreasing $\gamma$  due to spatial smoothing introduced from the larger pixel size. Our simulations show that a choice of $\gamma \approx 0.34$ is optimal for this specific scene.

\begin{table}[htp!]
\centering
{\footnotesize
\begin{tabularx}{\linewidth}{m|s|s|s|s|s|s}\hline
$\gamma$&3.43&1.71&0.69&\textbf{0.34}&0.23&0.13\\\hline
RMSE [mm] & 1.422 & 0.956 &0.714 &\textbf{0.632} &0.642&0.671 \\ 
\hline
\end{tabularx}
}
\caption{The RMSE between measurements with different sampling parameters ($\gamma$) and the ground truth (Simulated results).}
\label{table:simu}
\end{table}

\section{Discussion and Conclusion}
In this paper, we presented a novel ToF imager concept with unprecedented lateral resolution and depth resolution for the full-field three-dimensional surface measurement of macroscopic objects. Our experiments demonstrate the ability to reach depth precision down to 35 $\mu m$, that are only fundamentally limited by the roughness of the measured surfaces~\cite{goodman1975statistical}. We demonstrated our principle by implementing three different sensor prototypes exploiting different pixel-architectures.

The \textit{lock-in camera} based ToF system enables electronic rejection of unmodulated light and provides a heterodyne gain. This is especially useful for robust 3D imaging implementations that must accommodate a variety of objects with widely different surface roughness and reflectivity and other imaging scenarios like~\cite{willomitzer2019high, willomitzer2019synthetic}. However, the current version of the lock-in camera suffers from its low sensor resolution of only $300 \times 300$ pixels, which makes it impractical to scan objects in a large field of view if a high lateral resolution of the object surface is desired.

In an effort to address the shortcoming of the lock-in based ToF sensor architecture, we showed that an implementation of CCD/CMOS based FPA camera techniques can significantly increase the lateral (pixel) resolution. In a first implementation, we used a \textit{flutter-shutter camera} to modulate per-pixel readout and detection in a fashion similar to the lock-in camera. Our 3D reconstructions demonstrate very high spatial resolution while still providing the same sub-mm depth precision far superior to conventional ToF-sensors. However, there are drawbacks as well: Unlike the lock-in camera, the flutter shutter camera is not able to suppress the unmodulated light, which leads to a significant decrease of the signal-background ratio if the power in both interferometer arms is not carefully matched. Therefore, the flutter shutter based ToF 3D sensor is best suited for objects with relatively homogeneous reflectivity.

In an effort to highlight the versatility of our 3D sensing approach, we demonstrated synthetic wavelength 3D sensing using only a conventional commodity CMOS sensor. The result is a ToF sensor with high resolution and sub-mm precision that offers significant advantages in compactness and low-cost relative to the current generation of commercially available CWAM ToF sensors. However, this approach still faces further challenges before it can receive widespread adoption since temporal resolution and dynamic range are currently much worse than commercially available CWAM ToF sensors.   

Among the three depth sensor concepts presented in this work, we conclude that the most promising sensor implementation is based on the \textit{flutter-shutter camera}, due to its high sensor resolution and the relatively short time required to capture a 3D model. In the future, we plan to improve our flutter-shutter implementation and demonstrate high resolution 3D measurements of dynamic scenes. While 3D video cameras have already been introduced using structured light with comparable performance~\cite{Willo3DCam17}, we believe that a video frame rate variant of our ToF imager implementations will provide a compact and inexpensive solution for high resolution, sub-mm depth precision 3D video acquisition that will be of significant use for a large number of commercial and industrial applications.  

\section*{Funding}
Defense Advanced Research Projects Agency (DARPA) REVEAL (HR0011-16-C-0028); National Science Foundation (NSF) CAREER award (IIS-1453192).

\section*{Disclosures}
The authors declare no conflicts of interest.


\bibliography{micro3D}

\bibliographyfullrefs{micro3D}

\end{document}